 \documentclass[12pt]{article}
 \usepackage{graphicx}
 \usepackage{amssymb}
 \usepackage{amsmath}
 \usepackage{amsfonts}
 \usepackage{revsymb}
 \newcommand{\be}{\begin{equation}}
\newcommand{\ee}{\end{equation}}
\newcommand{\bea}{\begin{eqnarray}}
\newcommand{\eea}{\end{eqnarray}}
\newcommand{\bean}{\begin{eqnarray*}}
\newcommand{\eean}{\end{eqnarray*}}

\newcommand{\ra}{\rightarrow}

\begin{document}
\begin{titlepage}
\bigskip
\rightline{}

\bigskip\bigskip\bigskip\bigskip
\centerline {\Large \bf {Counting the Microstates of a Kerr Black Hole}}
\bigskip\bigskip
\bigskip\bigskip

\centerline{\large  Gary T. Horowitz and Matthew M. Roberts}
\bigskip\bigskip
\centerline{\em Department of Physics, UCSB, Santa Barbara, CA 93106}
\centerline{\em  gary@physics.ucsb.edu, matt@physics.ucsb.edu}
\bigskip\bigskip
\begin{abstract}
We show that an extremal Kerr black hole, appropriately lifted to M-theory, can be transformed to a Kaluza-Klein black hole in M-theory, or a D0-D6 charged black hole in string theory. Since all the microstates of the latter have recently been identified, one can exactly reproduce the entropy of an extremal Kerr black hole. We also show that the topology of the event horizon is not well defined in M-theory.
\end{abstract}
\end{titlepage}
\baselineskip=16pt
\setcounter{equation}{0}
\section{Introduction}

The entropy of an extremal (four dimensional) Kerr black hole is simply given in terms of its angular momentum $J$:
\be\label{kerrS}
S = 2\pi |J|
\ee
Since $J$ is naturally quantized, this formula is analogous to the entropy of supersymmetric black holes which is given in terms of their integer normalized charges. String theory has been very successful in exactly reproducing the entropy of a variety of supersymmetric black holes by counting appropriate microstates (see, e.g., reviews \cite{Peet:1997es,Mathur:2005ai} and references therein). We will show below that similar techniques can be applied to extremal Kerr to reproduce (\ref{kerrS}). 

In the early days of string theory, it was noted that the bound on perturbative string states $|J| \le M^2$ looked like the bound on Kerr black holes $|J| \le M^2$. However the first is really  $|J| \le \alpha' M^2$ while the second is $|J| \le G_4 M^2$, so an extremal Kerr black hole is not related to a maximally spinning string. (This is fortunate since a maximally spinning string does not have enough states to reproduce the entropy of a black hole.) Instead, we will see that the microstates of Kerr can be described in terms of D-branes.

It might seem strange that a neutral black hole, like Kerr, should be described in terms of charged objects such as D-branes. However, the same approach was used successfully last year to describe the entropy of certain neutral Kaluza-Klein black holes \cite{Emparan:2006it}. The idea is simply that some neutral black holes can be lifted to M-theory in such a way that the reduction to IIA string theory has both D0 and D6 charge. One can then count the number of D0-D6 bound states. This was shown to work for both static and rotating Kaluza-Klein black holes \cite{Emparan:2007en} as long as they had sufficient D0 and D6 charge after dimensional reduction.

In mapping Kerr to the class of Kaluza-Klein black holes whose entropy is understood, we will use standard tools such as T-duality and extrapolations between weak and strong coupling. The one key new ingredient
 is a transformation which allows us to exchange the angular momentum for a charge. Thus, in this context, angular momentum turns out to be equivalent to charge. This is not the first time that such an equivalence has been noted. For example, it was shown in \cite{Horowitz:1993jc} that T-duality on the BTZ black hole yields a charged black string in which the charge of the black string is directly related to the angular momentum of the black hole.
 
 In the course of our analysis, we
 will see that the topology of a black hole event horizon is not well defined in M-theory:  Equivalent descriptions of a black hole can can have different topology. Of course, given a horizon with topology $S^{2n+1}$, one can always view the sphere as a circle bundle over $CP^n$ and do T-duality along the circle. This changes the topology to $CP^n \times S^1$. However, in only one description is the horizon circle bigger than the string scale, and that is the one for which the supergravity description (and hence topology) is valid. We will present a different type of example where the supergravity description is valid for two topologically different black holes which nevertheless can be shown to be equivalent.
 
 In the next section, we briefly review the microstate counting for Kaluza-Klein black holes. In section three, we show how this counting can be applied to an extremal Kerr black hole and argue that horizon topology is not well defined. The final section has some concluding comments.  
 
\setcounter{equation}{0}  
 \section{Review of Kaluza-Klein microstates}
 
 Five dimensional neutral black holes, with translation invariance around the compact fifth direction, are described by four parameters. In terms of their reduction to four dimensions, these are the mass $M$, angular momentum $J$, and electric and magnetic charges $Q, P$. We are only interested in the extremal limit, in which $M$ is a function of the other parameters. This limit has qualitatively different behavior depending on
 whether $J$ is less than or greater than $|PQ|/G_4$.
 In this section we review the microstate counting of slow-rotating extremal Kaluza-Klein black holes in \cite{Emparan:2006it}. While this was extended to the fast-rotating case in \cite{Emparan:2007en}, to understand neutral Kerr black holes we only need to consider Kaluza-Klein black holes with $J=0$. In this case,
the mass and entropy are given by \cite{Larsen:1999pp}
\be
M_{bh}=\frac{(Q^{2/3}+P^{2/3})^{3/2}}{2 G_4}, \quad S_{bh}=2\pi\frac{|PQ|}{G_4}
\ee

 If we consider the Kaluza-Klein black hole as a five dimensional solution, multiplying by a constant (square) $T^6$ (with volume $(2\pi)^6V_6$) gives us a vacuum solution to M-theory. Reducing to IIA string theory by treating the Kaluza-Klein circle as the M-theory circle, we get a black hole with D0 and D6 charge. While D0-D6 states, like our black hole, are nonsupersymmetric, there are quadratically stable nonsupersymmetric D0-D6 bound states \cite{Taylor:1997ay,Witten:2000mf}. Recall that the M-theory reduction yields $R=gl_s$ and that further $T^6$ compactification gives $G_4=g^2 l_s^8/8 V_6$. The charge quantization from Kaluza-Klein theory translates, in IIA language, to 
\be\label{QP}
Q=\frac{g l_s}{4(V_6/l_s^6)} N_0,\quad P=\frac{g l_s}{4}  N_6
\ee
where $N_0, N_6$ are integers representing the number of D0 and D6 branes. Note that in terms of these integers, the entropy becomes simply $S_{bh} = \pi N_0 N_6$. 

Now suppose $N_0 = N_6 = 4N$. If
 we consider the $T^6$ as a product of three $T^2$'s and T-dualize along one cycle of each $T^2$, we get a configuration of four stacks of D3-branes wrapping the diagonal cycles of the $T^2$'s. There are $N$ branes in each stack. If the D3-branes were wrapping the fundamental cycles instead, this configuration would be equivalent to a four charge black hole whose microscopic entropy is known to be $S=  2\pi N^2$. Since this is independent of the moduli of $T^6$, it seems clear that the entropy is associated with the common intersection point of the branes. So when we rotate the branes to wrap the diagonals, the only change in the entropy is that there are now eight intersection points on $T^6$ (two on each $T^2$).
 Thus the entropy continues to agree: $S_{branes} = 8(2\pi N^2) = \pi N_0 N_6 = S_{bh}$. 
 
 More generally, if $N_0=4k^3N,\;N_6=4l^3N$,\footnote{One can properly describe the system in terms of intersecting three-branes only when $N_0$ and $N_6$ are of this form.} the dual system has four stacks of  branes each wrapping  the  cycles
 of the form $x_2=\pm kx_1 /l$ (see Fig. 1). 
 There are now $(2kl)^3$ intersection points in total,  again giving the correct entropy
\be
S_{branes}=(2kl)^3\times 2\pi N^2=\pi N_0N_6=S_{bh}.
\ee
 The mass of the D3-branes, which is proportional to their volume, also agrees with the mass of the black hole,
\be
M_{branes}=\frac{4N(k^2+l^2)^{3/2}V_6^{1/2}}{g l_s^4}=M_{bh}.
\ee

\begin{figure}[htp]
\centering
\includegraphics{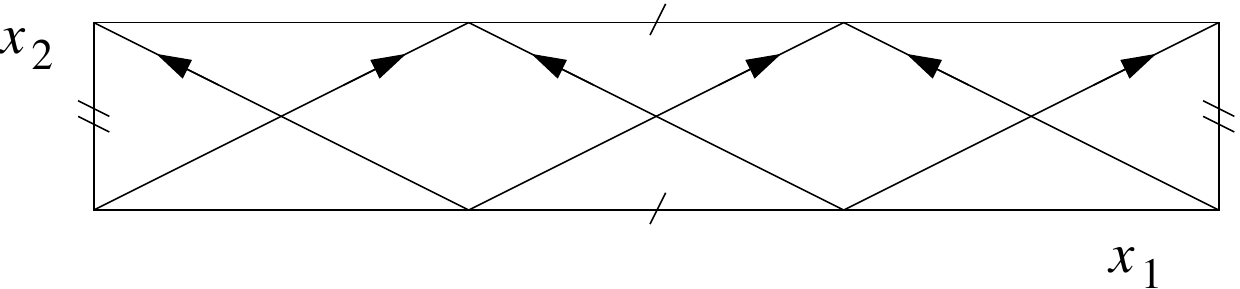} 
\caption{\label{fig:unequal}The branes wrap a rational direction $k/l$ of the torus (in the figure, $k=3$, $l=1$), so there are $2kl$ intersection points on each $T^2$. }
\end{figure}

 Since each intersection point contains four stacks with $N$ branes, and the microscopic counting is valid only for large charges, we require $N\gg 1$. 
While the restriction to specific forms of $N_0$ and $N_6$ seems constraining, it is worth noting that we can obtain any value of $P/Q$ we like by varying $V_6$ at fixed $N$, $k$, $l$.

\setcounter{equation}{0}  
 \section{Mapping Kerr  to a Kaluza-Klein black hole}
 
 We now show how to map an extremal Kerr black hole with angular momentum $J'$ into a {\it nonrotating} Kaluza-Klein black hole with large $N_0, N_6$ whose entropy was counted above.
 We  proceed in three steps which we will describe in terms of the quantum numbers of the extremal Kaluza-Klein black holes $ (N_0, N_6, J)$. 

{\it Step 1: Kerr $\ra (N_0 =0, N_6 = 1, J = J')$}

\noindent To begin, 
consider extremal Kerr cross a line, that is, a rotating black string. We now boost along the line (which of course does not change the local geometry) and compactify to a circle of radius $R$. The result is a rotating Kaluza-Klein black hole with electric charge. It is important to note that, for fixed angular momentum, the horizon area  does \emph{not} depend on the boost. This is easily seen from the form of the entropy for a rotating extremal black hole (this includes both the fast- and slow-rotating case) \cite{Emparan:2007en}
\be\label{extremal_entropy}S_{bh}=2\pi\sqrt{|N_0^2N_6^2/4-J^2|}\ee
while we have $N_6=0$. Specifically, we will boost to obtain $N_0=1$, i.e., we consider the minimum possible boost. Now, we again take the product of this solution with a $T^6$ to get  an M-theory solution, and by dimensionally reducing on the Kaluza-Klein circle get a IIA solution with one unit of D0 charge. We can then T-dualize along the entire $T^6$, and get a IIA solution with one unit of D6 charge. Now lifting this back to M-theory, we find a Kaluza-Klein black hole with one unit of \emph{magnetic} charge.

An important consequence of  these transformations is that the topology of the horizon in M-theory changes. Let us suppose that the $T^6$ is string scale (so it does not change size under T-duality) and only count macroscopic dimensions.  At weak coupling, the black hole has horizon topology $S^2$ and either D0 or D6 charge in the two equivalent descriptions. At strong coupling, the original black hole has topology $S^2\times S^1$ while the dual one has  $S^3$.  This is because, with $N_6=1$, the M-theory circle combines with the $S^2$ in the base to form a $S^3$. Since T-duality relates equivalent descriptions,  we see that the topology of the horizon of a black hole is not well defined in M-theory. Under this equivalence, a graviton probe of the horizon topology of one black hole maps into a Kaluza-Klein monopole probe of the other.

{\it Step 2: $(N_0 =0, N_6 = 1, J = J') \ra ( N_0 = 2J', N_6 =1 , J=0)$}
 
\noindent A rotating Kaluza-Klein black hole with one unit of magnetic charge can be thought of as a black hole sitting on the tip of a Taub-NUT space. For large enough $R$, the black hole looks like a five dimensional Myers-Perry black hole \cite{Myers:1986un}. The five dimensional angular momenta $J_{1,2}$ in the two orthogonal planes are related to $N_0, N_6, J$ by \cite{Emparan:2007en}
\be 
J_{1,2}=\frac{N_0 N_6}{2}\pm J
\ee
Since $N_0 =0$, our solution has $J_1=- J_2$.  A simple reflection will change the sign of $J_2$. However, a black hole with $J_1=+J_2$ corresponds to one with $J=0$ and $N_0$ nonzero. In other words, by reflecting the black hole before gluing it into the Taub-NUT, we exchange the four dimensional angular momentum with Kaluza-Klein circle momentum\footnote{This interesting fact has been noticed independently by R. Emparan.} (see Fig. 2). Since the reflection is clearly a discrete symmetry of the Myers-Perry solution, this transformation does not change the black hole. 

It is important to note that such a reflection is really only valid when $N_6=1$, and it is only for one magnetic charge that, in the large $R$ limit, the transformation above is an exact symmetry. Otherwise, our space does not become asymptotically flat when $R$ is large, but instead the asymptotic angular structure and the horizon topology are both $S^3/\mathbb{Z}_{N_6}$. 

\begin{figure}[htp]
\centering
\includegraphics{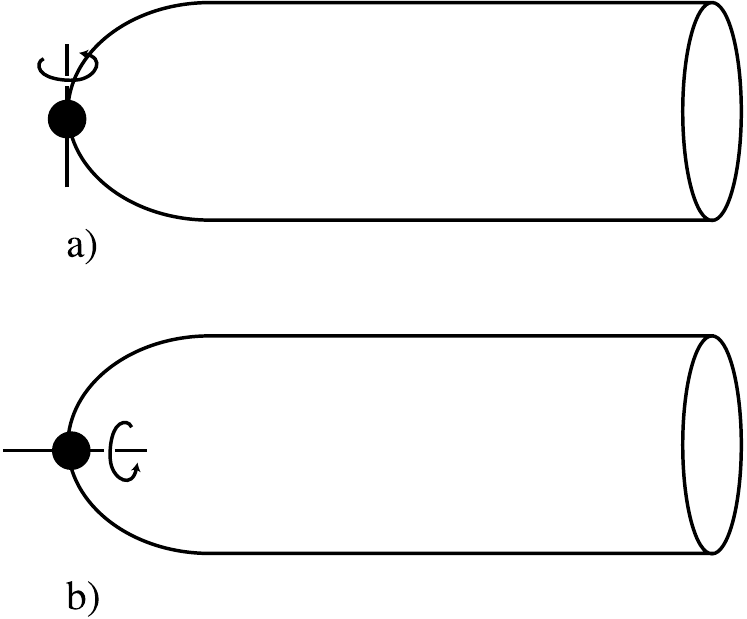} 
\caption{\label{fig:reflection} By taking $R$ large, the geometry becomes a Myers-Perry black hole at the tip of Taub-NUT. A simple reflection now changes configuration (a) with $N_0 = 0$ and $J\ne 0$ into (b) with $N_0 \ne 0$ and $J=0$. Although it appears that the black hole has been rotated by $90^o$, this is just an artifact of the projection down to two dimensions.}
\end{figure}

In the  limit of large $R$, the mass is invariant under the reflection. However, as we decrease $R$ the mass can change. This is easily seen when $R$ is small as the contribution to $M$ from $N_6=1$ is then negligible. Before the reflection, the black hole mass is like extremal Kerr  with $M=\sqrt{J/G_4}=\sqrt{J R V_6(2\pi)^7/ G_{11}}$, whereas afterwards it is an extremal electrically charged black hole with $M=N_0/R$. Not only do these scale differently with $R$,  the reflection symmetry translates angular momentum into electric charge as $N_0^{(new)}=2J^{(old)}$.

We must of course be careful about taking the $R\rightarrow \infty$ limit, as  this is the strong coupling limit of string theory. However, as our Kaluza-Klein black holes and their Myers-Perry limits possess an $SL(2,\mathbb{R})\times U(1)$ near-horizon symmetry \cite{Kunduri:2007vf}, there is an attractor mechanism which allows us to count microstates at weak coupling and extrapolate to strong \cite{Dabholkar:2006tb,Astefanesei:2006sy}. In fact, we can also use the argument \cite{Emparan:2007en} that as  there are flat directions in the dilaton's effective potential \cite{Astefanesei:2006dd}, only the entropy is attracted to the fixed weak-coupling value, and the mass is not guaranteed to be fixed, which clearly it is not. It is interesting to note that it is only in the \emph{strong} coupling limit that the masses do agree.

{\it Step 3:  $( N_0 = 2J', N_6 =1 , J=0) \ra (N_0, N_6$} large, $J=0)$

\noindent We have now transformed our Kerr black hole to a Kaluza-Klein black hole with $N_0$ large and $N_6 =1$. As previously described, our understanding of D0-D6 microstates is in the T-dual intersecting D3-brane picture. For this to be applicable, we require both $N_0$ and $N_6$ to be large.  To achieve this, we first T-dualize on the entire $T^6$ to obtain a solution with $N_6 = 2J'$  and $N_0=1$.   Geometrically, the $N_6$ charge corresponds to a quotient of the $S^3$  by identifying points along the Hopf fiber. If $K$ divides $N_6$, we can pass to a K-fold covering space in which we unwrap the Hopf fiber $K$ times. In taking the covering space, we want to keep the local geometry fixed, i.e., the supergravity parameters $Q,P$  are fixed as well as the eleven dimensional Planck length $l_p$. Since $l_p= g^{1/3} l_s$,  $R = g^{2/3} l_p$ so increasing $R$ by a factor of $K$ increases $g$ by $K^{3/2}$ and decreases $l_s$ by $K^{1/2}$.  From (\ref{QP}) it follows that $ N_0 \rightarrow K^2, \ N_6 \rightarrow N_6/K$.   The entropy, $S=\pi N_0 N_6$, increases by $K$ as expected since the horizon area is $K$ times larger. The entropy of the black hole in the covering space can now be reproduced exactly as shown in \cite{Emparan:2006it}. Since the covering space  geometrically is just $K$ copies of the  black hole, the original black hole has $S = \pi N_0 N_6 = 2\pi J'$ which indeed agrees with the entropy of the Kerr black hole we started with.

As further justification for this argument, we would like to see that in passing to the covering space, the microstates can be divided into $K$ identical, independent Hilbert spaces.  Starting with $N_0 = N_6 = 4N$ with $N=KL$, the $K$-fold cover has $N_0 = 4LK^3, \ N_6= 4L$. There are now $(2K)^3$ intersection points of the D3-branes, each giving rise to identical, independent Hilbert spaces.  So there is no difficulty in dividing them into $K$ groups. 

Strictly speaking, the counting in section two requires $N_0, N_6$ to be of the form $N_0 = 4k^3 N, \ N_6 = 4l^3 N$. This restricts the angular momentum of the original Kerr black hole. One possibility is to take $J' = 4 n^3$ for some large integer $n$: Setting  $N= n^2$ and $K=2n$, we have $N_0 = K^2 = 4N$ and $N_6 = 2 J' / K = 4N$. Of course, the integer $J'$ for any macroscopic black hole is enormous, and one can always find an integer $n$ such that  $J' \approx 4 n^3$. (To be precise, given a large integer $J$, there is an integer $n$ such that $(J-4n^3)/J < 3/n$.)

\setcounter{equation}{0}  
\section{Comments}
We have shown that one can reproduce the entropy of an extremal Kerr black hole by counting microstates in string theory. This was achieved by mapping the Kerr black hole into a class of Kaluza-Klein black holes whose entropy was recently counted. The map uses several transformations which are commonly used when discussing the entropy of supersymmetric black holes, such as dualities and extrapolations between weak and strong coupling. We have also included a discrete isometry of the black hole. This apparently innocuous transformation allows one to transform the angular momentum into a charge. 

Our first step was to consider the  product of the Kerr black hole and $S^1$, and add a small boost along the circle. While the black hole entropy does not depend on the boost, we do not have an independent argument that the number of microstates does not depend on the boost. Strictly speaking, we have counted the microstates of a rotating black string with one unit of momentum. However, for a macroscopic amount of angular momentum, the four dimensional reduction of this black string  is essentially indistinguishable from a standard Kerr black hole.

As in the original Kaluza-Klein case, we still must rely on the dual D3 description to count microstates. Counting the microstates in the D0-D6 system directly is very difficult, as it would require an understanding the moduli space of nonsupersymmetric D0-D6 states. One possible method would be to understand instanton-like field configurations in the D6 world-volume effective field theory, a six dimensional Euclidean $U(N_6)$ super Yang-Mills theory\footnote{The configurations we are looking for are ones satisfying $\int Tr F=\int Tr F\wedge F=0$, $\int Tr F\wedge F\wedge F \propto N_0$, and are at least perturbatively stable.}. However, it is worth pointing out that while we do not have a D3 picture for arbitrary values of $N_0$ and $N_6$, it is shown in \cite{Witten:2000mf} that the D0-D6 system is stable while not supersymmetric.

Several open questions remain. As mentioned above, the entropy counting only applies to extremal Kerr with certain values of $J$. Extending this to other values of $J$ is related to extending the Kaluza-Klein entropy counting to general  values of $N_0, N_6$. One could also ask about higher dimensional rotating black holes. The fact that all five dimensional extremal Myers-Perry black holes with nonzero area can be obtained as large $R$ limits of rotating Kaluza-Klein black holes means that their entropy can be counted in this manner.  Understanding the entropy of six and higher dimensional rotating black holes remains open. Of course, one would also like to go beyond the extremal limit to near extremal or nonextremal black holes. Finally, the counting of states for Kerr that is described here depends on a toroidal compactification. There should be analogous ways to count microstates for other compactifications. 

\vskip 1cm
\centerline{\bf Acknowledgements}
It is a pleasure to thank D. Berenstein, H. Elvang, R. Emparan, S. Gukov and D. Marolf for discussions. This work was supported in part by NSF grant PHY-0555669.

 \end{document}